\documentclass{PoS}

\title{Fluctuation studies in STAR}

\ShortTitle{Fluctuation studies in STAR}

\author{\speaker{Supriya Das (for the STAR Collaboration)}\thanks{Present address: Gesellschaft für Schwerionenforschung mbH (GSI), Planck Str. 1, 64291 Darmstadt, Germany.}\\
        Variable Energy Cyclotron Centre\\
	1/AF, Bidhannagar, Kolkata 700064\\
	India.\\
        E-mail: \email{S.Das@gsi.de}}

\abstract{Study of event by event fluctuations of thermodynamic quantities offer us more insight about the hot and dense matter created in the relativistic heavy ion collisions. In this review the recent results on these studies carried out by the STAR collaboration are presented.}

\FullConference{The 2nd edition of the International Workshop --- Correlations and Fluctuations in Relativistic Nuclear Collisions --- \\
		 July 7-9 2006\\
		 Galileo Galilei Institute, Florence, Italy}

\def\ave#1{\langle {#1} \rangle}

\begin{document}

\section{Introduction}

Study of fluctuation and correlation have special importance while investigating the existence of possible phase transition in relativistic heavy ion collisions. Event by event fluctuations in the thermodynamic quantities provide important insight towards the physical properties of the dense nuclear matter produced in the collisions along with the knowledge of nature of quark-hadron phase transition \cite{heiselberg}. The enhanced fluctuation of energy density points to first order phase transition whereas a second order phase transition may result into divergence of specific heat. Study of event by event fluctuation in $\ave{p_{\rm T}}$, charged and neutral multiplicity correlation are tools those could be used to look for the nature and order of the phase transition. Another aspect of these studies is to search for the existence and location of the critical point on the quark-hadron phase diagram \cite{stephanov}. It is believed to observe large fluctuations of thermodynamic quantities in the vicinity of the critical point.

Now, the measured fluctuation from the relativistic heavy ion collisions may have contributions from many different sources. Firstly there are quantum fluctuations, which arise if the specific observable does not commute with the Hamiltonian of the system under consideration. But these fluctuations probably play a less significant role in heavy ion collisions. Secondly and most important in our case are the so called ``dynamical fluctuations'' reflecting the dynamics and response of the system under external forces. They help to characterize the bulk properties of the system like the instability arising at the phase boundary if the system undergoes a phase transition. Finally there are contributions from known sources including (a) geometrical (impact parameter, participant number, detector acceptance), (b) energy momentum and charge conservation, (c) anisotropic flow, (d) Bose-Einstein correlations, (e) resonance decays, (f) correlations from jets and minijets {\it etc}. These contributions are needed to be understood and subtracted in order to estimate the true extent of dynamical fluctuation that tell us about the properties of the system. One or more of the following methods are used to achieve this

\begin{itemize}

\item construction of mixed events which contain the known fluctuation contributions
\item simulation of known fluctuations using the physical phenomena those contribute to them
\item formulating proper quantity which is robust against contributions from known sources
 
\end{itemize}
  
With the advent of the recent experiments with relativistic heavy ion collisions many fascinating results are available for the understanding of the QGP phase transition in particular as well as the overall nature of the dense matter in general. In this review the results obtained from Solenoidal Tracker At RHIC (STAR) experiment will be discussed. 

\section{STAR experiment}

STAR experiment comprises of a Time Projection Chamber (TPC), as its central tracking detector, with a pseudorapidity coverage of $|\eta|$ <1.0 and azimuthal coverage $-\pi < \phi <\pi$. All results discussed here in this review are obtained from analysis of data taken by the TPC. The charged tracks are identified by using their relative energy loss in TPC. Collision centralities are measured by the number of charged particles with the pseudorapidity range $|\eta|$ < 0.5. Glauber Monte Carlo has been used to estimate the number of participants involved in the collisions for each class of centrality.

The data used for the results in this review are from Au+Au collisions at $\sqrt {s_{NN}}$ = 20, 62.4, 130 and 200 GeV.

\section{ $\ave{p_{\rm T}}$ correlation and fluctuation}

 $\ave{p_{\rm T}}$ of the particle produced in a collision corresponds to the temperature of the system. So the event by event fluctuation in  $\ave{p_{\rm T}}$ were proposed to estimate the fluctuation in temperature. But recent $\ave{p_{\rm T}}$ fluctuation studies reveal contributions from many other sources including flow and minijiets to it.  In STAR there are several analyses studying the event by event fluctuation/correlation of  $\ave{p_{\rm T}}$.

\begin{figure}[htbp]
\includegraphics[width=18pc]{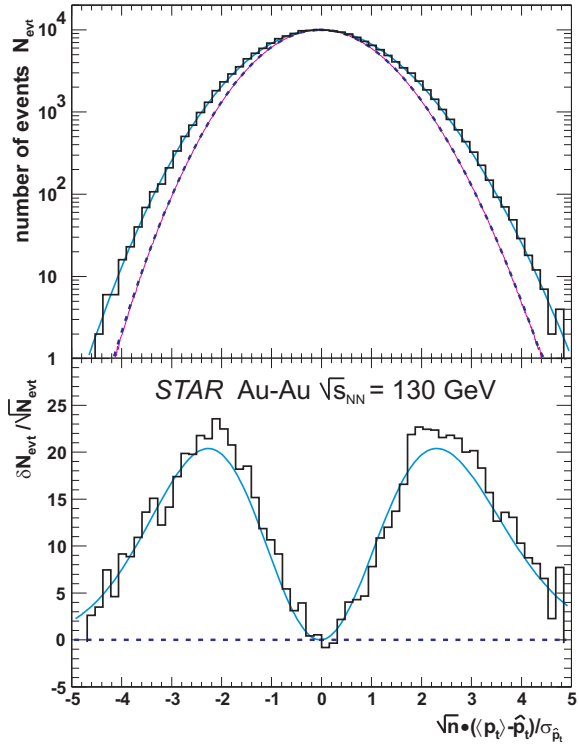}
\hspace{2pc}
\begin{minipage}[b]{15pc}\caption{\label{fig1_prc71}$\ave{p_{\rm T}}$ fluctuation. Upper panel: Event frequency distribution of $\ave{p_{\rm T}}$ fluctuation measure from data (histogram) compared to statistical references, such as, $\Gamma$ distribution (dashed curve), Monte Carlo reference (solid curve underlying $\Gamma$ distribution). The broadened distribution (solid curve underlying data) is obtained by raising it to a power calculated from the measured value of fluctuation from statistical analysis (see text). Lower panel: Difference in upper panel, between data and $\Gamma$ reference (histogram) and difference between $\Gamma$ reference and broadened distribution (solid curve).}
\end{minipage}
\end{figure}

\begin{figure}[htbp]
\includegraphics[width=16pc]{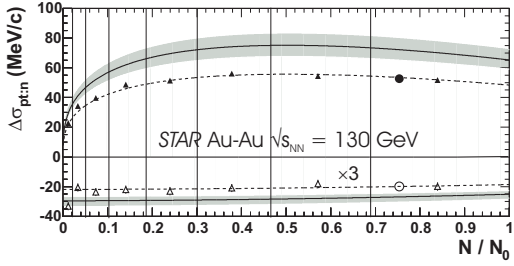}
\hspace{2pc}
\begin{minipage}[b]{16pc}\caption{\label{fig2_prc71}Centrality dependence of $\ave{p_{\rm T}}$ fluctuation. Charge independent (solid triangular points) and charge independent (open triangular points) difference. Parameterization (dashed curves) extrapolation of parameterizations to true primary particle number (solid curves), systematic uncertainties (color bands).}
%\vspace{5cm}
\end{minipage}

\end{figure}

Fig.~\ref{fig1_prc71} shows a 14\% deviation of event by event $\ave{p_{\rm T}}$ distribution from the statistical reference constituted (a) from $\Gamma$ function distribution and (b) from Monte Carlo simulation \cite{prc71} for central Au+Au collisions at $\sqrt {s_{NN}}$ = 130 GeV. The broadened $\Gamma$ function distribution is obtained by raising it to a power calculated from the measured value of fluctuation, $\Delta\sigma_{pt:n}$, from the statistical analysis. The agreement of this broadened distribution with the data histogram ensures the fact that the measured fluctuation excess is a property of most particles in most collisions, is not produced by a minority of outliers. The excess in width of data over the reference occurs rather smoothly on both sides of the peak. Absence of significant deviation in this rules out evidence of anomalous event classes, as might be expected from critical fluctuations. In Fig.~\ref{fig2_prc71} charge dependent and charge independent  $\ave{p_{\rm T}}$ difference factors are plotted in dependence of event centrality. This shows that the charge independent dynamical fluctuation is more than the charge independent ones and it varies smoothly and monotonically with collision centrality.  

\begin{figure}[h]
\begin{minipage}{18pc}
\includegraphics[width=18pc]{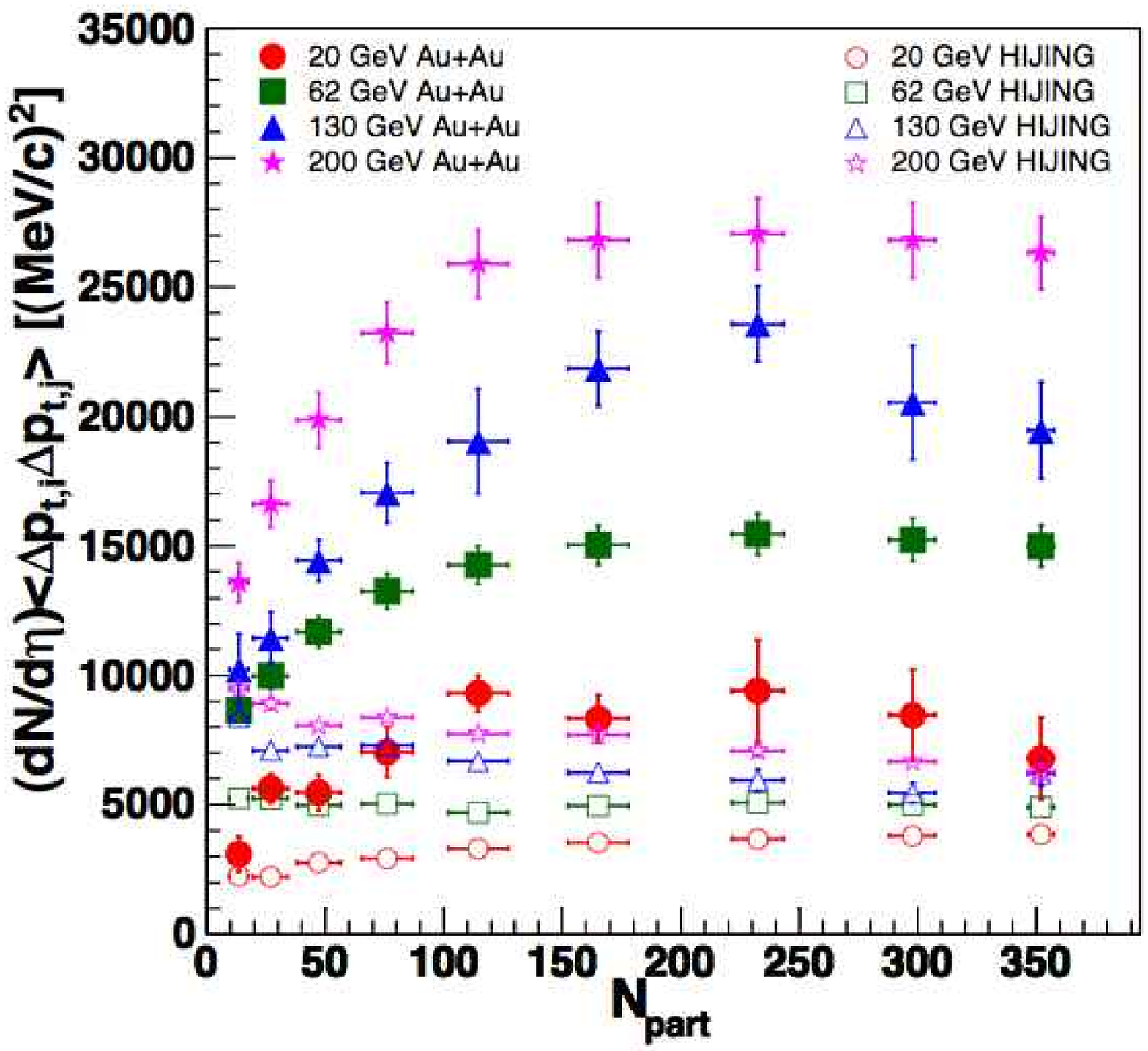}
\caption{\label{fig1_prc72}$(dN/d\eta)\langle \Delta p_{t,i}
\Delta p_{t,j} \rangle$ as a function of centrality and incident
energy for Au+Au collisions compared with HIJING results.}
\vspace{2pc}
\end{minipage}\hspace{2pc}
\begin{minipage}{16pc}
\includegraphics[width=16pc]{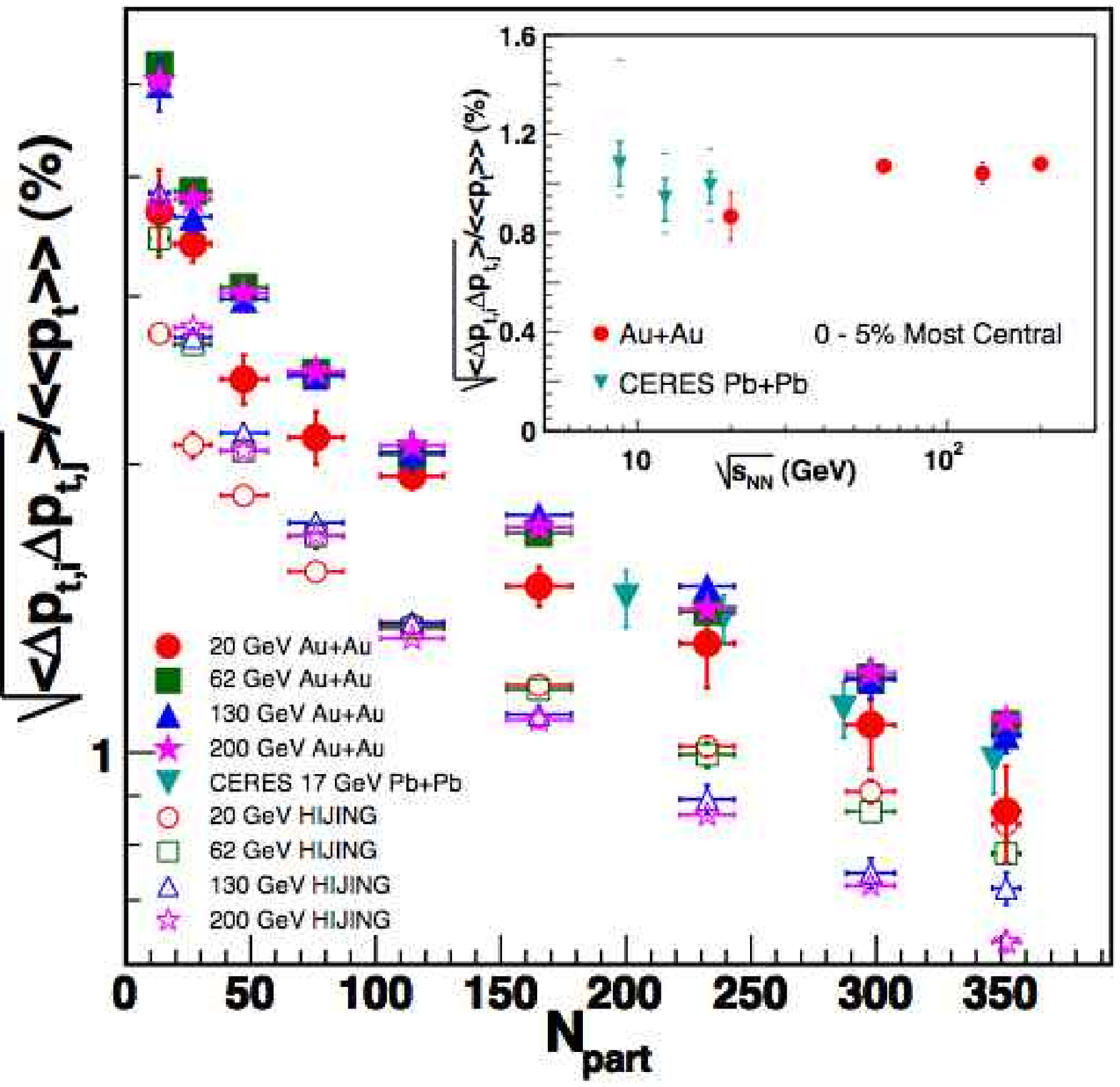}
\caption{\label{fig2_prc72}$\sqrt{\langle \Delta p_{t,i}
\Delta p_{t,j} \rangle}/\left\langle {\left\langle {p_t}
\right\rangle } \right\rangle$ as a function of centrality
and incident energy for Au+Au collisions compared with
HIJING results for corresponding systems.
The inset shows the excitation function for the most central bin.}
\end{minipage}
\end{figure}

Another analysis in STAR \cite{prc72} uses two particle correlation approach as a measure of dynamical fluctuation. The $\ave{p_{\rm T}}$ correlation show non-zero values for all collision energies available at RHIC.  Fig.~\ref{fig1_prc72} shows that for all collision energies the scaled value of correlation increases rapidly with centrality and then saturates for central collisions. To study the possible changes in the dynamical fluctuation with the change in the event average of $\ave{p_{\rm T}}$ ($\ave{\ave{p_{\rm T}}}$) with centrality and incident energy, the dynamical fluctuation scaled by $\ave{\ave{p_{\rm T}}}$ is plotted in Fig.~\ref{fig2_prc72} as a function of centrality as well as collision energies. Results from HIJING model calculation underpredict the measured values of correlation and do not predict the observed centrality dependence.

\vspace{0.5cm}

\begin{figure}[h]
\begin{tabular}{cc}

\begin{minipage}{10pc}
\includegraphics[keepaspectratio,width=1.65in]{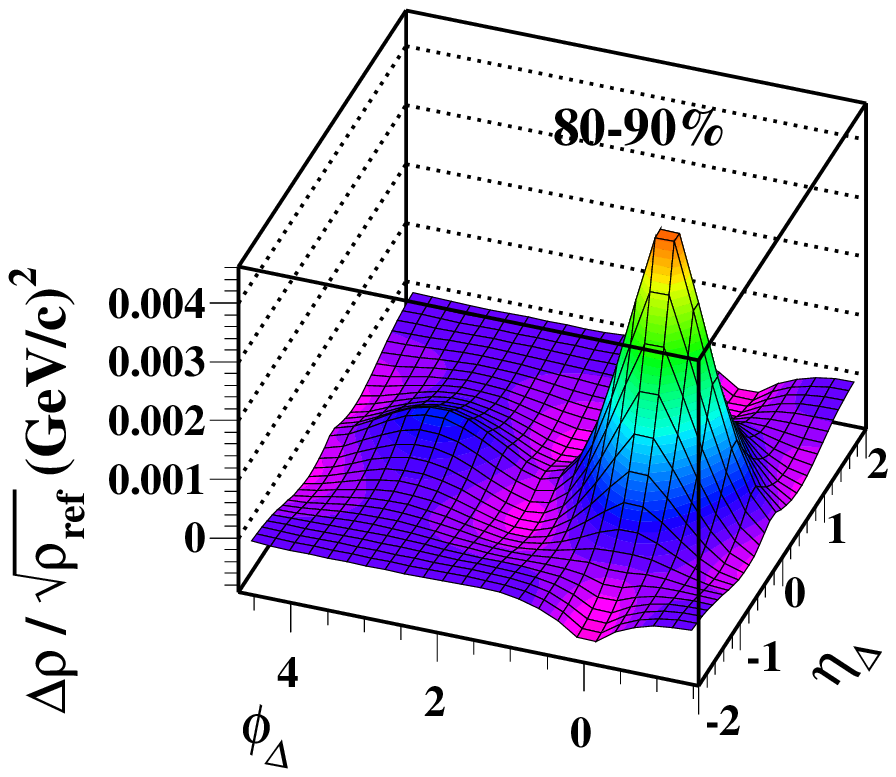}
\end{minipage} & \hspace{0.5pc}
\begin{minipage}{10pc}
\includegraphics[keepaspectratio,width=1.65in]{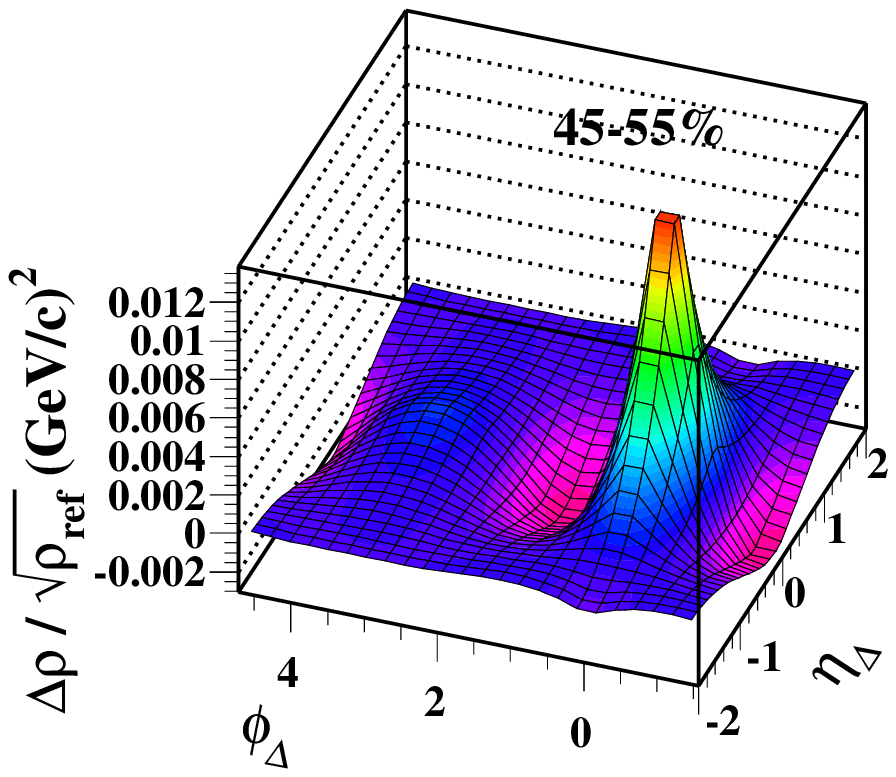}
\end{minipage}  \\ \hspace{0.5pc} 
\begin{minipage}{10pc}
\includegraphics[keepaspectratio,width=1.65in]{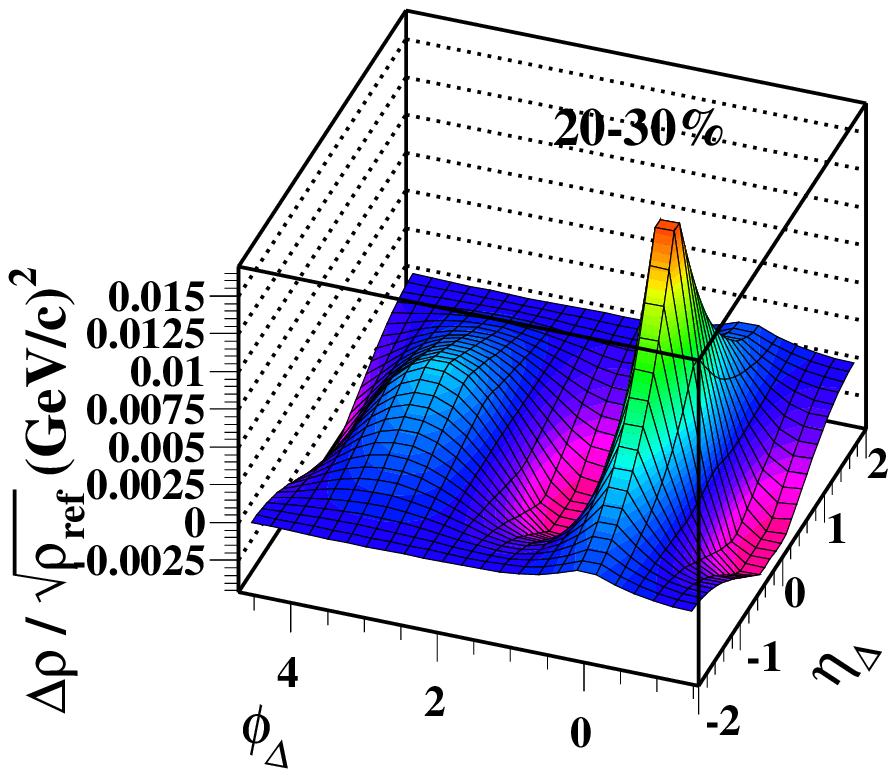}
\end{minipage} & \hspace{0.5pc}
\begin{minipage}{10pc}
\includegraphics[keepaspectratio,width=1.65in]{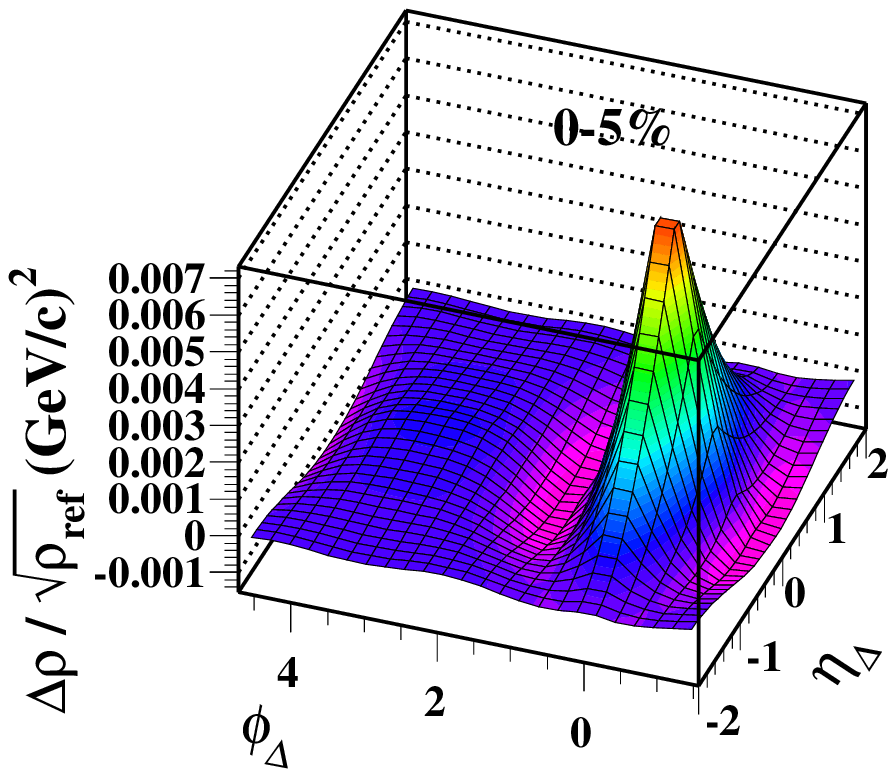}
\end{minipage}\\
\end{tabular}
\begin{minipage}[b]{12pc}\caption{Distributions of $\Delta \rho / \sqrt{\rho_{ref}}(\eta_\Delta,\phi_\Delta)$  for 80-90\% (upper left), 45-55\% (upper right),  20-30\% (lower left) and 0-5\% (lower right) of total cross section. Monopole (constant), dipole and quadrupole components have been subtracted.   \label{fig_jpg32}}
\end{minipage}
\end{figure}

There is another approach in STAR to study the scale dependence of $\ave{p_{\rm T}}$ correlations and associated angular autocorrelations \cite{jpg32}. In this approach the scale dependence of $\ave{p_{\rm T}}$  fluctuations in ($\eta,\phi$) is measured and the corresponding autocorrelations in difference variables ($\delta\eta, \delta\phi$) are obtained by inverting the $\ave{p_{\rm T}}$ distributions. These autocorrelations reveal interesting structures in phase space, which are attributed to production of minijets in relativistic heavy ion collisions. Fig.~\ref{fig_jpg32} shows the results from $\ave{p_{\rm T}}$ scale dependence for Au+Au collisions at $\sqrt {s_{NN}}$ = 200 GeV.

%-----------------------------------
%-----------------------------------

\begin{figure}[htbp]
\begin{center}
\vspace{0.5cm}
\includegraphics[scale=0.65]{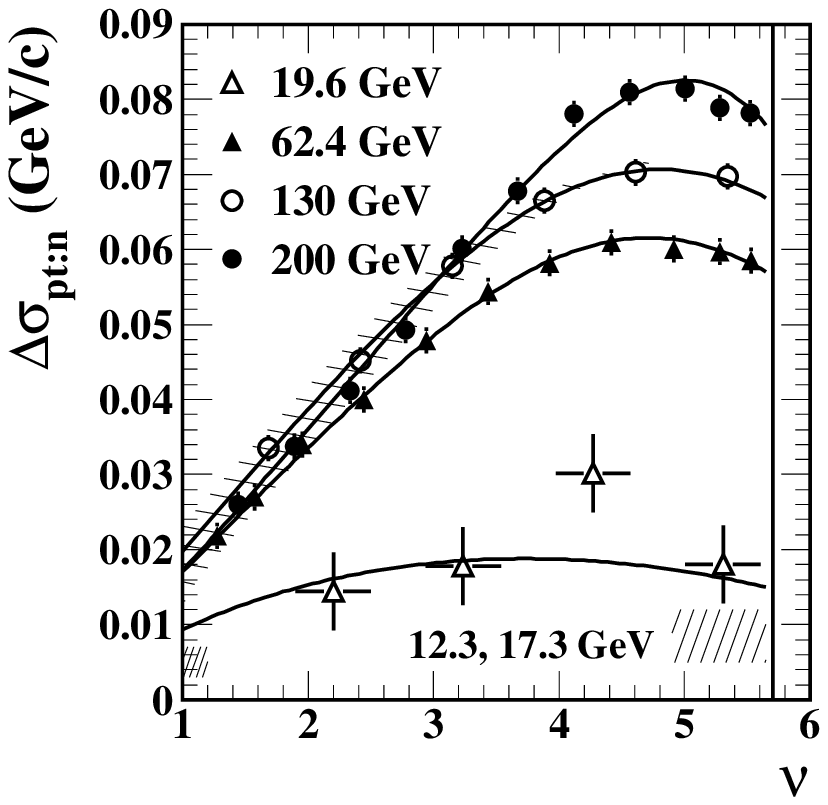}
\hspace{6pc}
\includegraphics[scale=0.65]{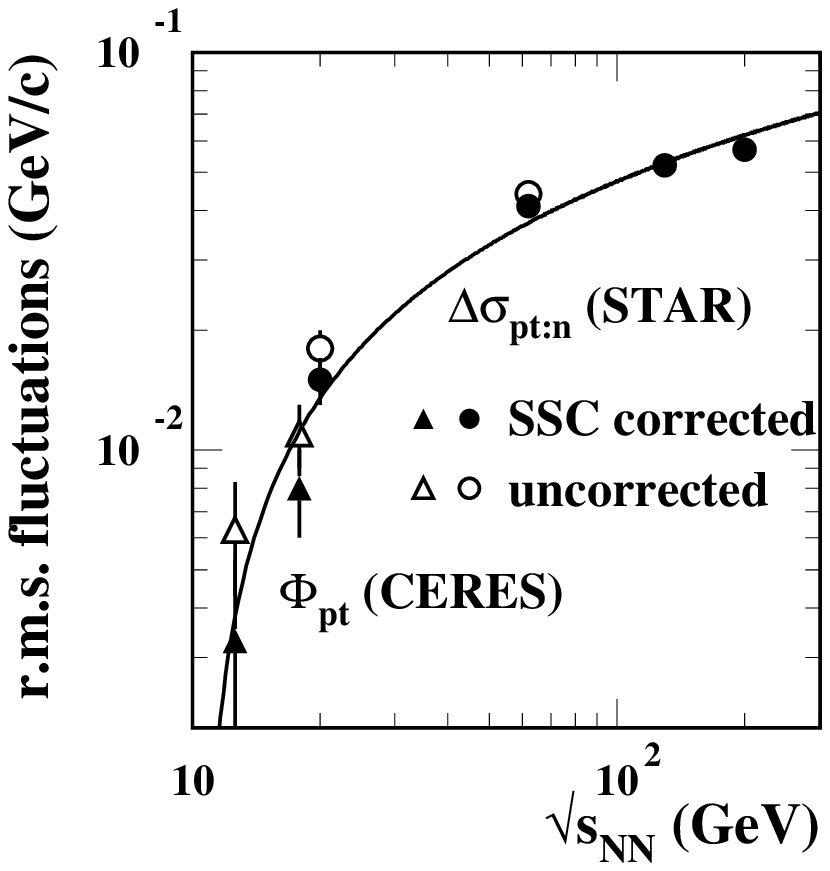}
\caption{Scale dependence of $\ave{p_{\rm T}}$ angular correlation, centrality dependence (left), energy dependence (right).}
\label{fig_estruct_edep}
\end{center}
\end{figure}

There has been a continuation of this approach by using the variance difference in $\ave{p_{\rm T}}$ angular correlation to determine the centrality and collision energy dependence of this quantity in STAR \cite{estruct_edep}. Fig.~\ref{fig_estruct_edep} shows the results from this analysis. The left panel shows the centrality dependence of $\ave{p_{\rm T}}$ fluctuation for all collision energies at RHIC. The centrality has been measured as $\nu \sim 2N_{bin}/N_{part}$ where $N_{bin}$ is average number of binary collisions and $N_{part}$ is the average number of participant nucleons. The vertical line at right estimates $\nu$ for most central collisions (b=0). The lines are drawn to guide the eye. In the right panel the energy dependence of $\ave{p_{\rm T}}$ fluctuation for most central collisions is plotted. The full and open symbols correspond to data with and without SSC (HBT, Coulomb) correction. The solid curve is proportional to $\ln (\sqrt {s_{NN}}/10)$.

\section{Net Charge fluctuation}

Fluctuation in conserved quantity like electric charge has been proposed to be a probe to the know about the QGP phase. There have been a number of theoretical investigations predicting the value of charge fluctuations for confined hadronic phase as well as deconfined QGP phase \cite{ahm,jk}. With the wide acceptance TPC of STAR the dynamic net charge fluctuations have been measured for Au+Au collisions for all available collisions energies. 

\begin{figure}[htbp]
\includegraphics[width=20pc]{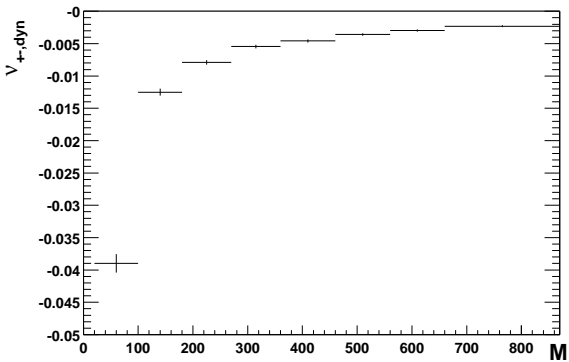}
\hspace{2pc}
\begin{minipage}[b]{12pc}\caption{Centrality dependence of dynamical net charge fluctuation for Au+Au collisions at $\sqrt {s_{NN}}$ = 130 GeV.}
\label{fig1_prc68}
\end{minipage}
\end{figure}

\begin{figure}[htbp]
\begin{minipage}{16pc}
\includegraphics[width=16pc]{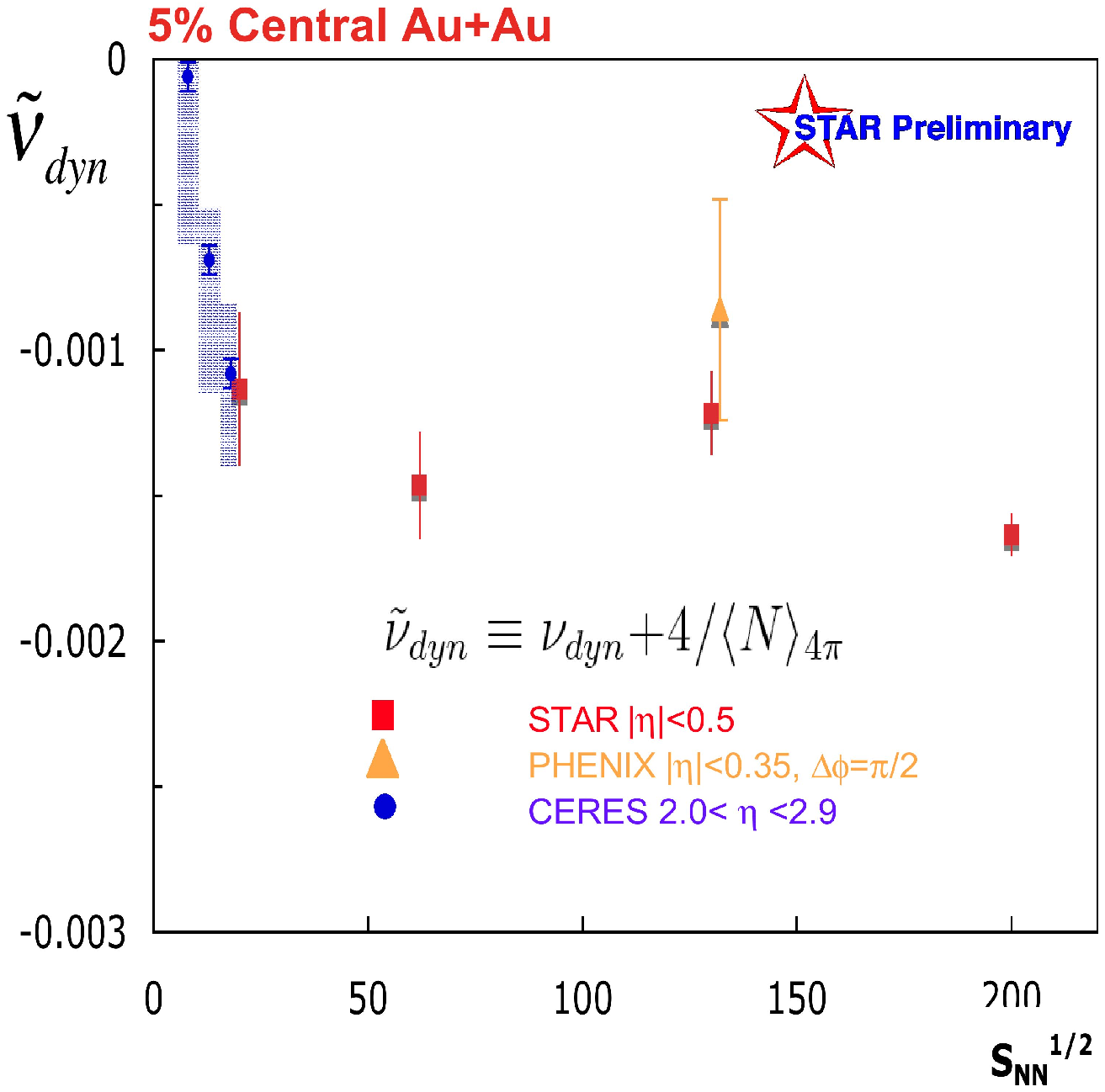}
\caption{\label{fig_netcharge_energy}Excitation function for dynamical net charge fluctuation. Data from SPS also plotted for better comparison.}
\vspace{2pc}
\end{minipage}\hspace{4pc}
\begin{minipage}{16pc}
\includegraphics[width=16pc]{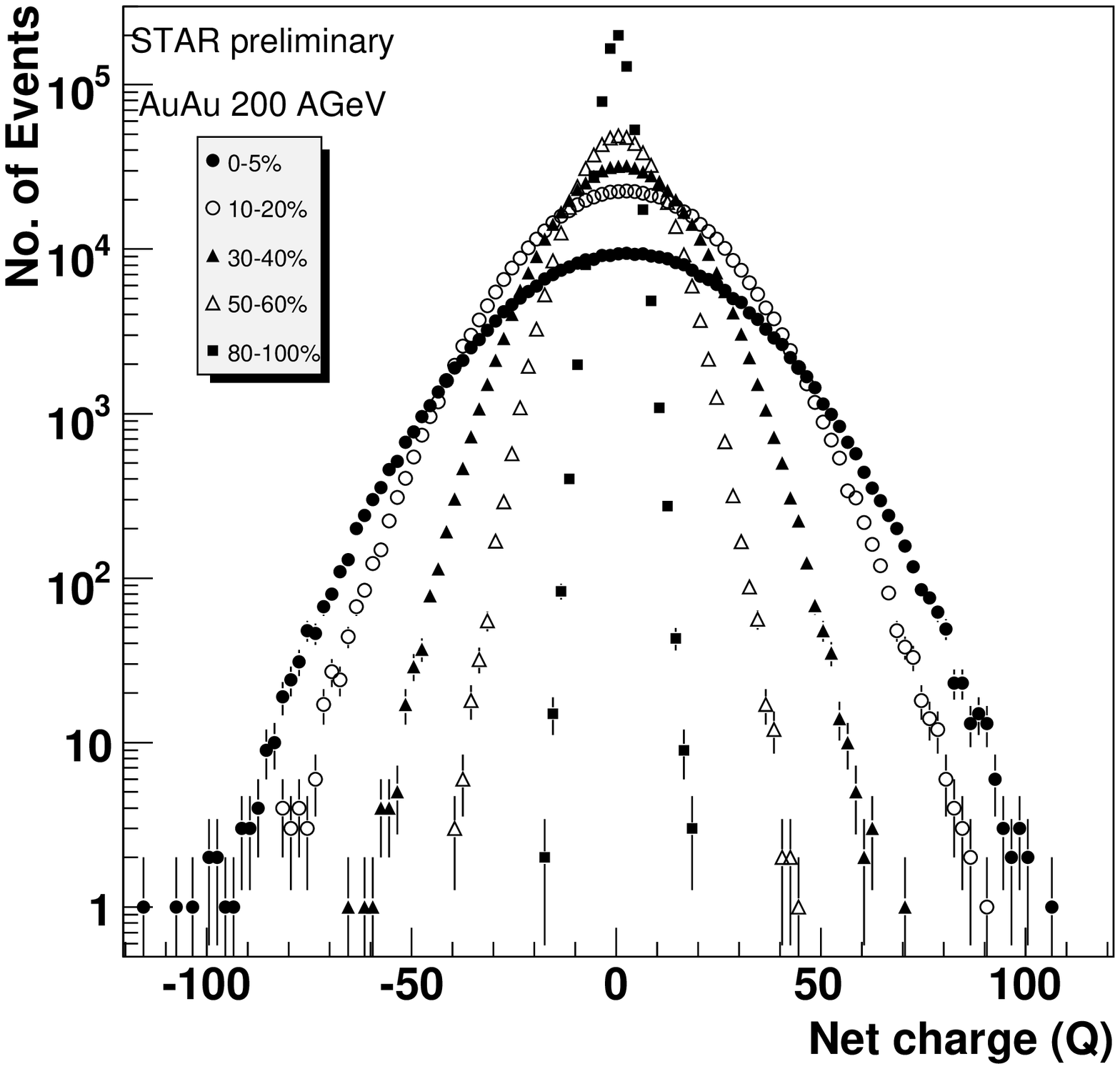}
\caption{\label{fig_sqm06_net_charge}Net charge distributions from charged particles with $p_{\rm T}$ below 1 GeV/c for Au+Au collisions at $\sqrt {s_{NN}}$ = 200 GeV}
\vspace{1cm}
\end{minipage}
\end{figure}

Fig.~\ref{fig1_prc68} shows the centrality dependence of the dynamical net charge fluctuation measured through a correlation function approach \cite{prc66} for Au+Au collisions at $\sqrt {s_{NN}}$ = 130 GeV \cite{prc68}. There is a strong centrality dependence of the dynamical fluctuation and the two particle correlation might have been modified in central collisions relative to peripheral ones. The measured value of fluctuation is closer to the prediction for resonance gas. Fig.~\ref{fig_netcharge_energy} shows that the dynamical net charge fluctuation is invariant of collision energies starting from top SPS energies \cite{netcharge_energy}.

Recent lattice calculations \cite{lattice1,lattice2} predict an enhanced charge fluctuations in the hadronic phase where as suppression of the same at the high temperature phase of QGP. Moreover, these calculations observe prominent structures in the higher order moments of net charge distributions for temperature close to the transition temperature. Fig.~\ref{fig_sqm06_net_charge} shows net charge distributions from charged particles with $p_{\rm T}$ below 1 GeV/c for Au+Au collisions at $\sqrt {s_{NN}}$ = 200 GeV \cite{sqm06_net_charge} for different centralities. Analysis of higher moments of net charge distribution is underway to extract new information.

\section{Particle ratio fluctuation}

Large difference in the relative production of kaons to pions has long been predicted due to the difference in free enthalpy of the hadronic and QGP phase \cite{strange_enhance}. Two of the most interesting and widely discussed experimental observations in the area of event by event fluctuation in relativistic heavy ion collisions are the strangeness fluctuation measured in terms of ratio of Kaons over Pions at SPS energies by NA49 Collaboration \cite{na49_prl} as well as the collision energy dependence of the measured fluctuation \cite{gunthar_qm04}.

\begin{figure}[htbp]
\begin{minipage}{18pc}
\includegraphics[width=18pc]{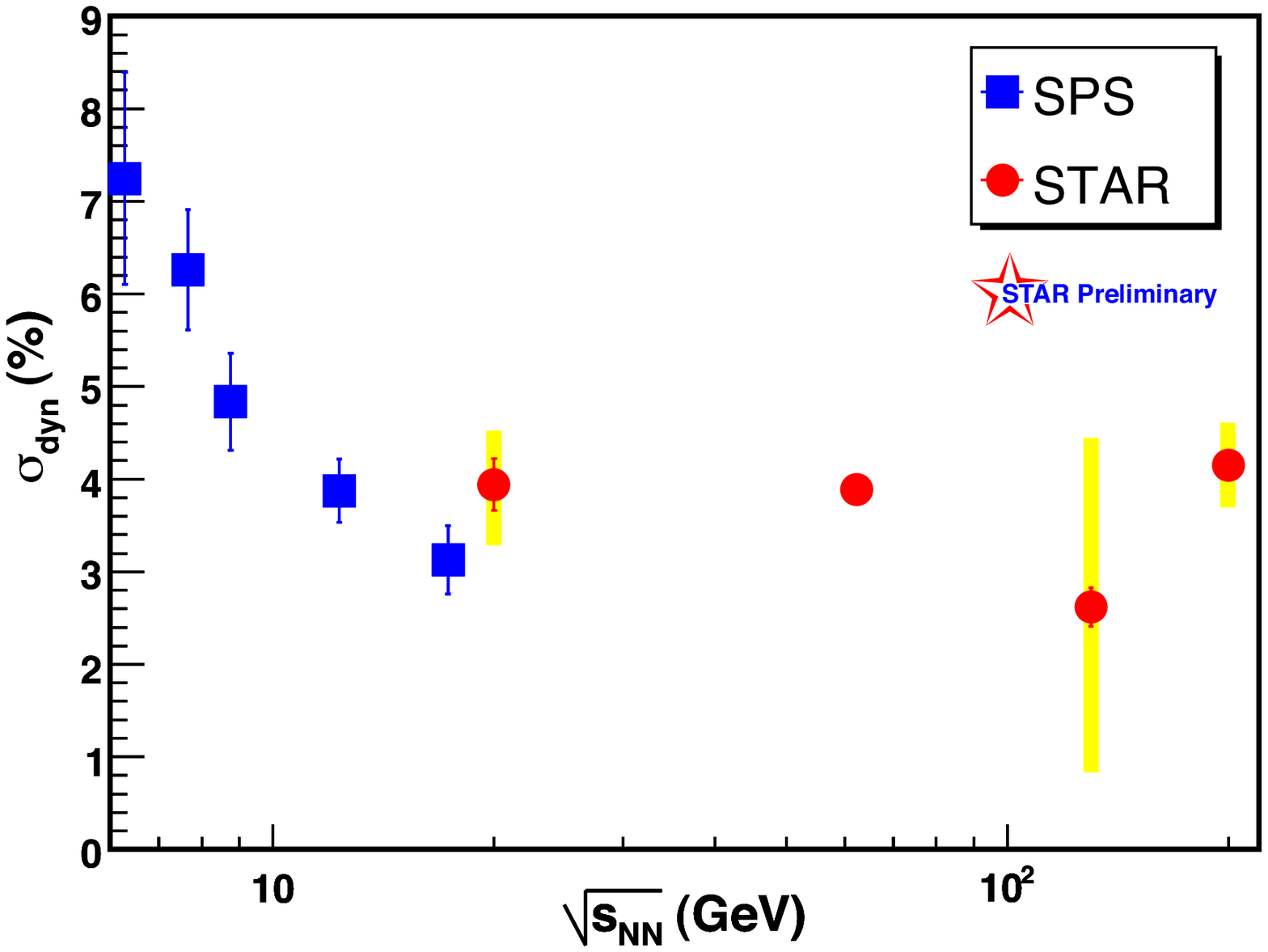}
\caption{\label{fig1_sqm06_kpi}Excitation function for the dynamical fluctuation in $K/\pi$ ratio. The data from SPS are plotted for comparison.}
\end{minipage}\hspace{2pc}
\begin{minipage}{16pc}
\includegraphics[scale=0.35]{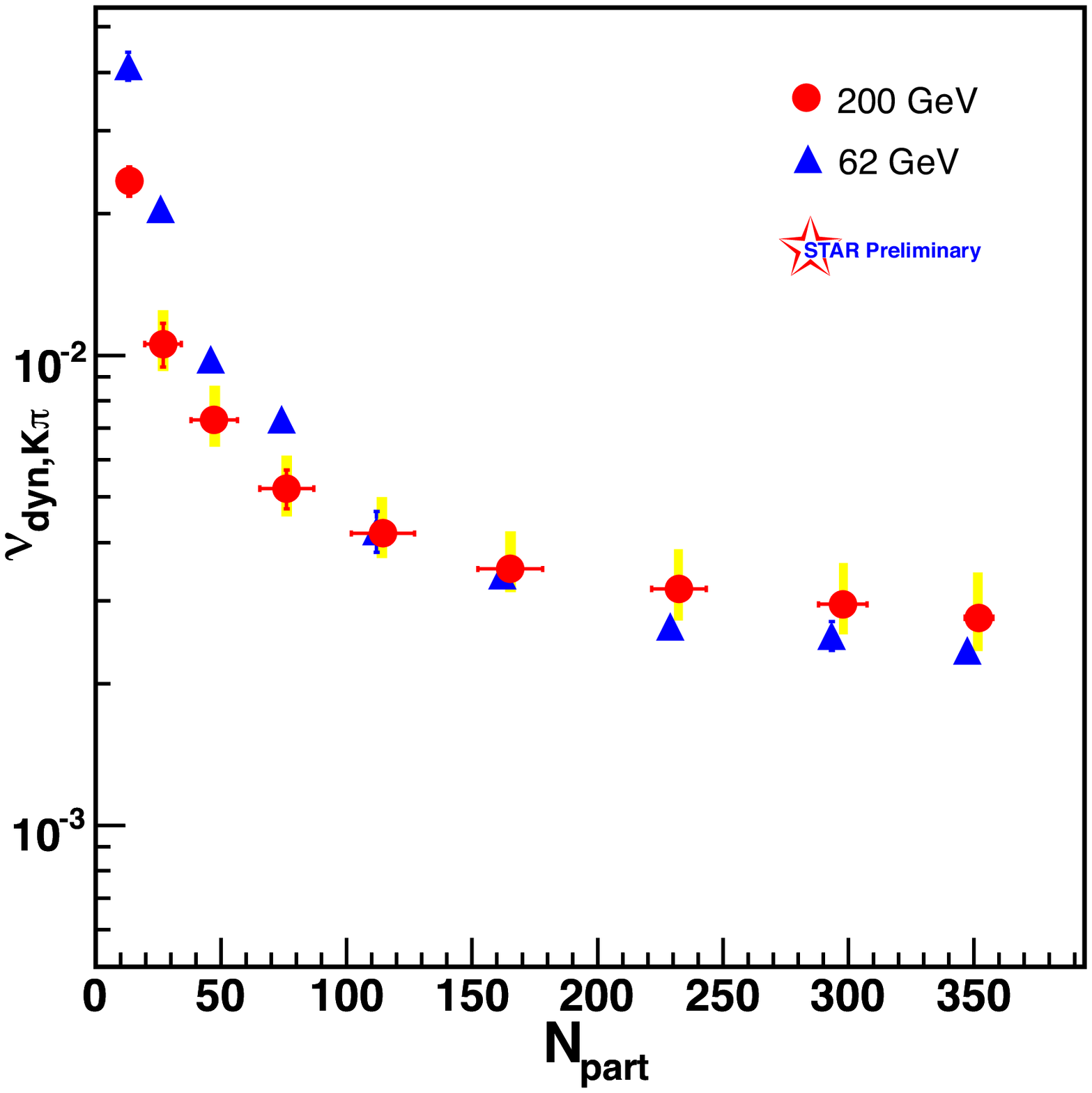}
\caption{\label{fig2_sqm06_kpi}Centrality dependence of dynamical fluctuation.}
\end{minipage}
\end{figure}

STAR experiment has also measured the event by event fluctuation in $K/\pi$ ratio in Au+Au collisions in all available collision energies at RHIC \cite{sqm06_kpi}. The kaons and pions are identified from the relative energy deposition by the charged tracks in TPC. The number of Kaons as well as pions are calculated event by event and the ratio is obtained. The dynamical fluctuation is extracted by subtracting the statistical reference, constructed by mixed event technique, from the measured value of fluctuation. Fig.~\ref{fig1_sqm06_kpi} shows the excitation function of dynamical fluctuation. The data points for NA49 have also been shown in this plot for comparison. Fluctuation in $K/\pi$ remains independent of the collision energies available at RHIC. The centrality dependence of dynamical fluctuation in Au+Au collisions at $\sqrt {s_{NN}}$ = 62.4 and 200 GeV are also studied using a correlation function approach similar to the net charge fluctuation analysis. Fig.~\ref{fig2_sqm06_kpi} shows that the measured value of dynamical fluctuation decreases with increasing collision centrality. This dilution of fluctuation could be attributed towards increasing production of correlated pairs from resonances as the collision becomes more and more central.

\section{Balance function}

One interesting probe to know about the history of hadronization after the relativistic heavy ion collisions is balance function \cite{bf_theory}. In the collision when the particles and anti particles are produced in pairs, they are correlated in coordinate space. If the hadronization occurs early, the charge - anti charge pairs are expected to be separated in rapidity due to expansion of the fireball as well as the re-scattering in the strongly interacting matter. On the other hand if the pair is produced in later time the the correlation is stronger. As the width of the correlation reflects the time of hadronization, this could signal to creation of high temperature QGP phase.

\begin{figure}[htbp]
\includegraphics[scale=0.45]{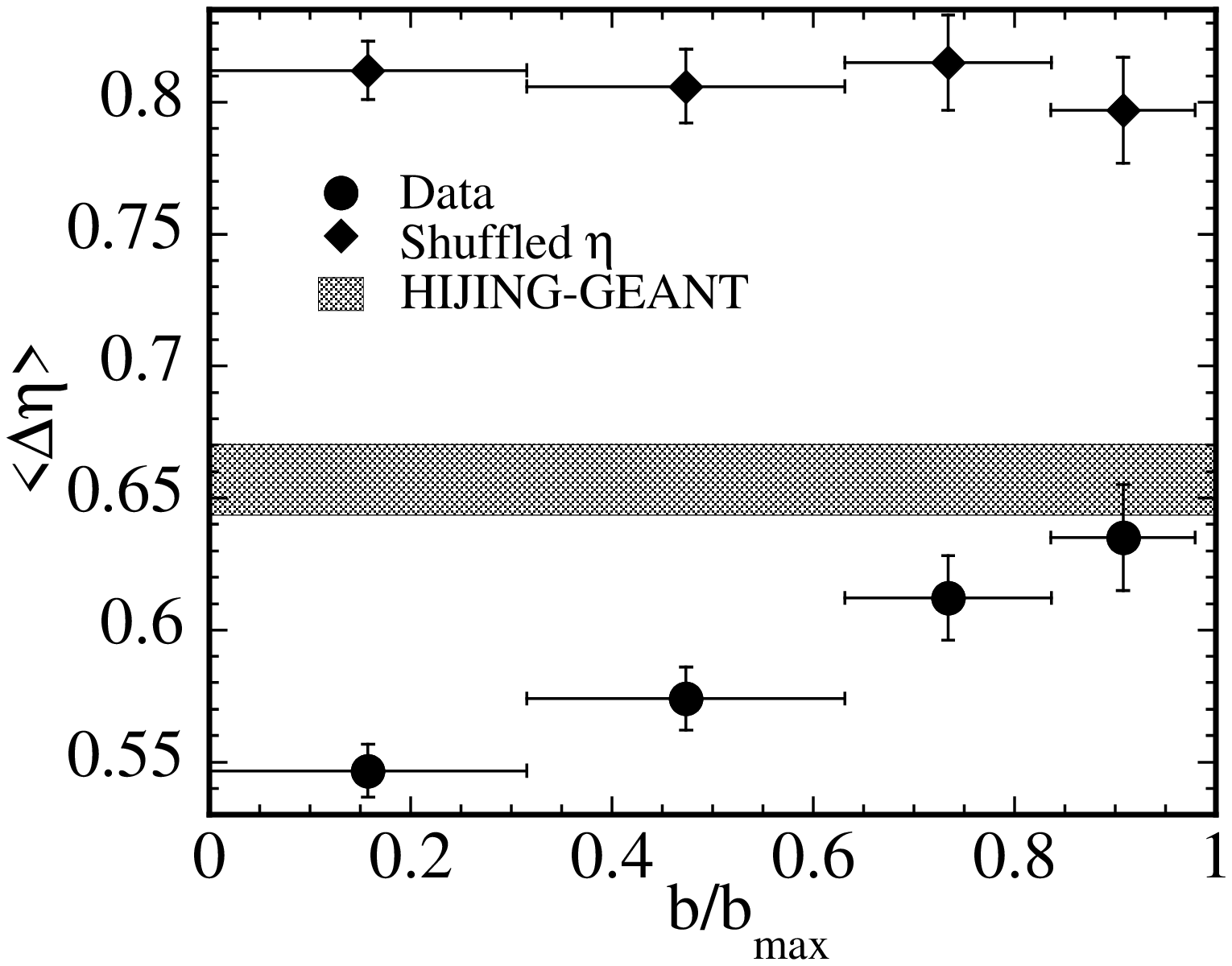}
\hspace{2pc}
\includegraphics[scale=0.45]{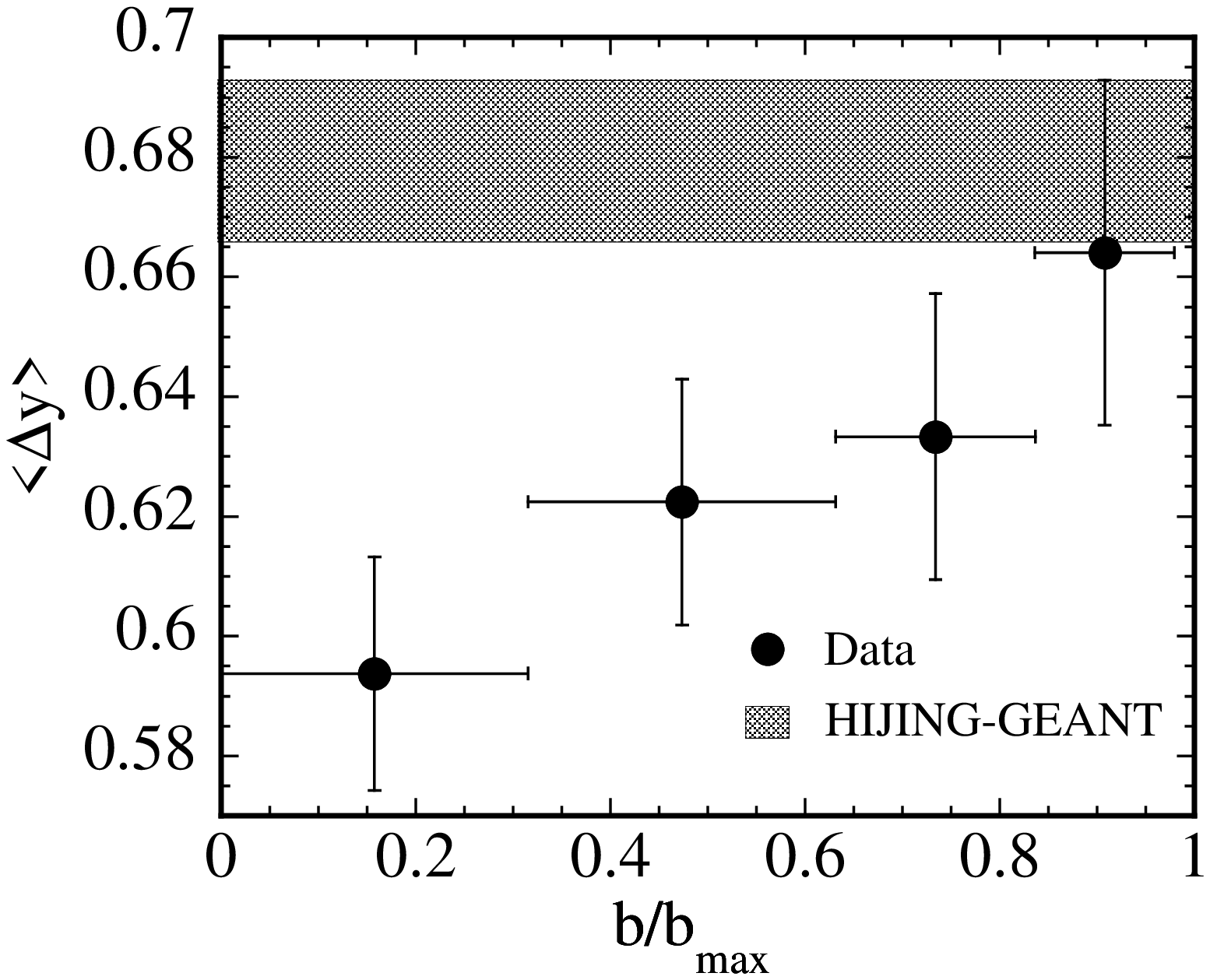}
\caption{Width of the balance function for unidentified charged particles (left) and identified pions (right). The result from HIJING calculation in both cases are shown in the shaded band.}
\label{fig_prl90}
\end{figure}

The width of the balance function for unidentified charged particles as well as identified pions (Fig.~\ref{fig_prl90}) in Au+Au collisions at $\sqrt {s_{NN}}$ = 130 GeV are observed to be narrower in case of more central collisions \cite{prl90}. In spite of sensitivity of this observation to effects like flow, resonance production and diffusion, it also indicates towards possibility of late hadronization when the collision is more violent. The width predicted from HIJING model calculation is consistent only with results obtained in peripheral collisions.

%\begin{figure}[h]
%\begin{minipage}{18pc}
%\includegraphics[width=18pc]{fig1_prl90.eps}
%\caption{\label{fig1_prl90}The width of the BF for identified charged pions as a function of normalized impact parameter. The result from HIJING calculation are shown in the shaded band.}
%\vspace{2pc}
%\end{minipage}\hspace{2pc}
%\begin{minipage}{16pc}
%\includegraphics[scale=0.36]{fig2_sqm06_kpi.eps}
%\caption{\label{fig2_sqm06_kpi}Centrality dependence of dynamical fluctuation.}
%\vspace{0.5cm}
%\end{minipage}
%\end{figure}

\section{Forward backward multiplicity correlation}

\begin{figure}[htbp]
\begin{center}
\includegraphics[scale=0.7]{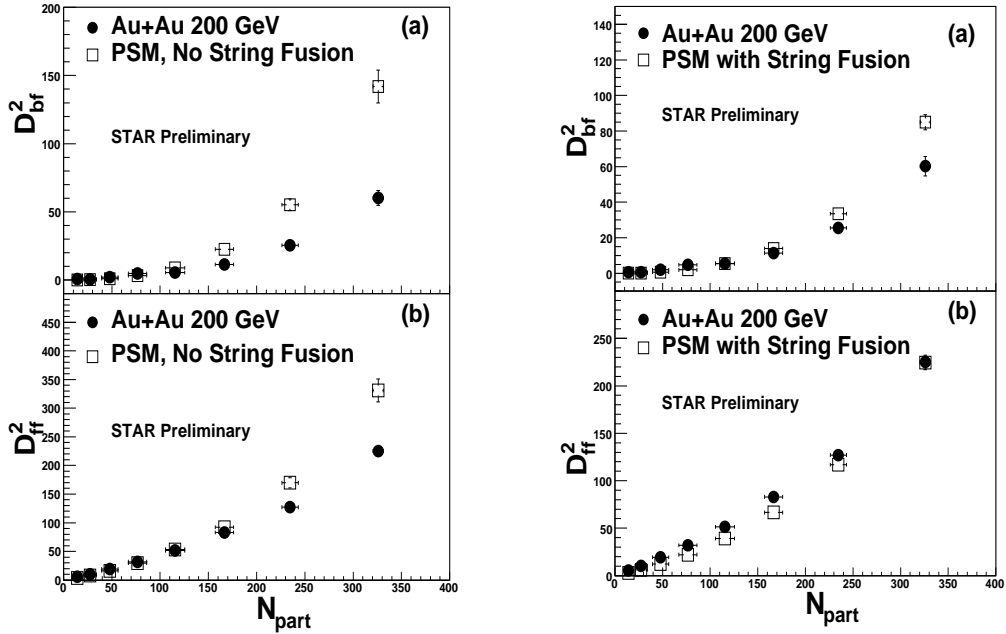}
\end{center}
\caption{$D^2_{bf}$, $D^2_{ff}$ as a function of centrality for Au+Au collisions at $\sqrt {s_{NN}}$ = 200 GeV. Data are compared with PSM calculations with and without string fusion.}
\label{fig_nuex_lrc}
\end{figure}

Short and long range (in rapidity) multiplicity correlations are predicted as a signature of string fusion. From the relationship between the multiplicities in the forward and backward pseudorapidity regions, a correlation coefficient could be defined as,

\begin{equation}
b=\frac{<N_fN_b> - <N_f><N_b>}{<N^2_f> - <N_f>^2} = \frac{D^2_{bf}}{D^2_{ff}}
\end{equation}
where $N_f$ and $N_b$ are the multiplicity in the forward and backward region respectively and $D^2_{bf}$, $D^2_{ff}$ are the {\it backward-forward} and  {\it forward-forward} dispersions.
When strings fuse a reduction of the long range forward backward correlation is expected. Fig.~\ref{fig_nuex_lrc} shows the centrality dependence of {\it backward-forward} and  {\it forward-forward} dispersions measured in Au+Au collisions at $\sqrt {s_{NN}}$ = 200 GeV. These results are obtained by considering the charged particles within $0.8 < |\eta| < 1.0$ \cite{nuex_lrc}. The measured values are also compared to those from Parton String Model (PSM) calculation with and without string fusion.  The reduction in the long range multiplicity correlation, $D^2_{bf}$, even compared to the collective PSM, is a possible indication that additional collective effect is present in central Au+Au collision at $\sqrt{s_{NN}}$ = 200 GeV. 

\section{Summary and outlook}

The wide acceptance of the STAR detector both in pseudorapidity ($|\eta|<1.0$) and azimuth ($-\pi < \phi <\pi$) allowed us to study event by event fluctuations in various observables like $\ave{p_{\rm T}}$, net charge, $K/\pi$ ratio as well as correlation analyses with balance function and forward-backward multiplicity correlation. Results from $\ave{p_{\rm T}}$ fluctuation analysis show substantial excess in fluctuation, over the statistical reference, which scales smoothly and monotonically with centrality. $\ave{p_{\rm T}}$ analysis in finer ($\eta, \phi$) sub-space shows interesting structures which could be attributed to production of minijets at RHIC. The centrality dependence of multiplicity scaled value of $\ave{p_{\rm T}}$ correlation may show signs of effects such as thermalization, the onset of jet suppression. Net charge fluctuation is greater than the predicted value from charge conservation limit but closer to resonance gas picture. Amount of dynamical net charge fluctuation remain essentially unaltered as a function of collision energies available at RHIC. Results from analysis of higher moments of net charge are promising but needs more work to obtain a more comprehensive understanding. STAR obtains the first results from event by event fluctuation in $K/\pi$ ratio at RHIC. This fluctuation is positive for all collision energies available at RHIC, however it remains practically independent of the collision energy. The dynamical fluctuation in $K/\pi$ ratio decreases as the collision become more and more central. Results from balance function analysis indicates towards a delayed hadronization after the heavy ion collisions at RHIC. A reduction in long range multiplicity correlation, compared to model calculation, suggests presence of additional collective phenomena in central heavy ion collisions.

One of the most important aspect of the fluctuation studies in relativistic heavy ion collisions is to search for the location of the critical point on the quark-hadron phase diagram. This could be achieved by scanning the QCD phase diagram in terms of temperature and baryonic chemical potential. A program has recently been proposed to run RHIC at low energies to probe the region of low temperature and rather high baryonic chemical potential for this purpose. The STAR experiment has a shinning future in this program. With the addition of a full barrel Time Of Flight (TOF) detector, STAR will increase the momentum range of identification of K and $\pi$ to play a key role in studying strangeness fluctuation near the phase boundary. It is also expected to have more insights from the energy dependence of balance function at lower energies.   

With the continued developments of new analysis techniques as well as upgrades of the detector subsystems, STAR will continue to provide more and more comprehensive knowledge about the quark hadron phase transition in relativistic heavy ion collision through correlation fluctuation studies.

\end{document}